\documentclass[12pt,preprint]{aastex}

\newcommand{\Teff}{\mbox{$T_{\rm eff}$}}

\newcommand{\Msun}{\mbox{$\cal M_{\odot}$}}

\newcommand{\fuvmag}{\mbox{\it FUV~}}
\newcommand{\nuvmag}{\mbox{\it NUV~}}
\shortauthors{Bianchi et al.}
\shorttitle{Object Classification  from GALEX and SDSS surveys}
\begin{document}
\title{Classification and Characterization of Objects from the GALEX and SDSS surveys}
\author{
Luciana Bianchi\altaffilmark{1}, 
Mark Seibert\altaffilmark{2}, 
Wei Zheng\altaffilmark{1}, 
David A. Thilker\altaffilmark{1},
Peter G. Friedman\altaffilmark{2},
Ted K. Wyder\altaffilmark{2},
Jose Donas\altaffilmark{4},
Tom A. Barlow\altaffilmark{2}, 
Yong-Ik Byun\altaffilmark{3}, 
Karl Forster\altaffilmark{2},
Timothy M. Heckman\altaffilmark{1}, Patrick N. Jelinsky\altaffilmark{6},
Young-Wook  Lee\altaffilmark{3}, Barry F. Madore\altaffilmark{7},
Roger F.  Malina\altaffilmark{4},
D. Christopher Martin\altaffilmark{2}, Bruno Milliard\altaffilmark{4},
Patrick  Morrissey\altaffilmark{2}, Susan G. Neff\altaffilmark{9},
R. Michael Rich\altaffilmark{10}, David Schiminovich\altaffilmark{2},
Oswald H. W.  Siegmund\altaffilmark{6}, Todd Small\altaffilmark{2},
Alex S. Szalay\altaffilmark{1}, Barry Y. Welsh\altaffilmark{6}
}
\altaffiltext{1}{
Deptartment of Phys.\& Astron., Johns Hopkins
University,3400 N.Charles St., Baltimore, MD21218
(bianchi,zheng,dthilker,heckman,szalay@pha.jhu.edu)}
\altaffiltext{2}{California~Inst. of Technology,MC405-47,1200 E.California Blvd, Pasadena, CA91125 
(mseibert,friedman,tab,krl,cmartin,patrick,ds,tas,wyder@srl.caltech.edu)}
\altaffiltext{3}{Center for Space Astrophysics, Yonsei University, Seoul
120-749, Korea
(byun,ywlee@obs.yonsei.ac.kr)}
\altaffiltext{4}{Laboratoire d'Astrophysique de
Marseille, BP8, Traverse du Siphon, 13376 Marseille Cedex 12,FR
(denis.burgarella,roger.malina,bruno.milliard@oamp.fr)}
\altaffiltext{6}{Space Sciences Laboratory, University of California at
Berkeley, 601 Campbell Hall, Berkeley, CA 94720
(patj,ossy,bwelsh@ssl.berkeley.edu)}
\altaffiltext{7}{Observatories of the Carnegie Institution of Washington,
813 Santa Barbara St., Pasadena, CA 91101
(barry@ipac.caltech.edu)}
\altaffiltext{9}{Laboratory for Astronomy and Solar Physics, NASA Goddard
Space Flight Center, Greenbelt, MD 20771
(neff@stars.gsfc.nasa.gov)}
\altaffiltext{10}{Department of Physics and Astronomy, University of
California, Los Angeles, CA 90095
(rmr@astro.ucla.edu)}

\begin{abstract}

We use the
 GALEX (Galaxy Evolution Explorer) Medium Imaging Survey (MIS) and
All-Sky Imaging Survey (AIS)  data available in the first internal release,
 matched to the SDSS catalogs in the overlapping regions,
to classify objects  by comparing the multi-band photometry to 
model colors.  We show an example of the advantage of such 
broad wavelength coverage (GALEX 
far-UV and near-UV, SDSS {\it ugriz}) in classifying objects and augmenting the
existing samples and catalogs.  From the MIS [AIS] sample over an area of 75 [92] 
square degrees, we select a total of  1736  [222] QSO candidates at redshift $<$ 2,
significantly extending the number of fainter candidates, and moderately increasing the
number of bright objects in the SDSS list of  spectroscopically confirmed QSO.
 Numerous  hot stellar objects are also revealed by the 
UV colors, as expected. 
\end{abstract}





\section{Introduction}
\label{sintro}

The GALEX (Galaxy Evolution Explorer) mission is providing the first
UV sky surveys with wide area coverage and deep sensitivity,
in two UV bands, far-UV (FUV) and near-UV (NUV). The major
science objectives and characteristics of GALEX, and of the surveys,
 are described by Martin et al (2004).

The portions of the Medium Imaging Survey (MIS) and All Sky Imaging Survey (AIS)
completed at the time of the Internal Release 0.2
(Morrissey et al. 2004) substantially overlap
with the Sloan Digital Sky Survey (SDSS)
 imaging and spectroscopic optical surveys, 
for a total 72 and 92 square degrees respectively.
 We present in this work examples of  analyses 
of the color-color digrams which illustrate the
advantages of combining two UV bands and five optical bands, 
in classifying and characterizing objects of certain types, and
within certain parameter ranges.  We classify the matched GALEX and
SDSS sources by comparing  their observed colors to model colors.

This paper complements the work by Seibert et al. (2004),
which presents a statistical analysis of the magnitude distributions
in the correlated GALEX and SDSS sample, and of the characteristics of objects
with previously available classification within the sample. 
Several other papers in this issue derive statistical properties
of galaxies from this sample (alone or correlated to catalogs
in other wavelength bands), e.g. Buat et al. (2004), Budavari et al. (2004), 
Arnouts et al. (2004), Treyer et al. (2004), Yi et al. (2004), Salim et al. (2004), Rich et al. (2004). 
Some of these works 
basically adopt in the definition
of the sample the star/galaxy  classification provided by the
SDSS pipeline, which define as ``star'' a source appearing point-like
at the SDSS resolution, 
and as ``galaxy'' an extended source.
Also, these works are mainly focused on color-magnitude diagrams and 
luminosity functions, while the present analysis is focused on 
the interpretation of color-color diagrams.

\section{The sample and the  Analysis}
The GALEX 
on-orbit performance is described by Morrissey et al. (2004). 
GALEX FUV (1350--1750~\AA) and NUV (1750--2750~\AA) imaging
is used in this work, with optical imaging from the SDSS
in the {\it ugriz} bands.

 In the current GALEX Internal Data Release, IR0.2, the matched
GALEX+SDSS surveys cover a non-contiguous area of the sky (excluding overlaps)
of 75 square degrees (MIS) and 92 square degrees (AIS).
The MIS has exposure times varying between 1000 and 1700 sec.,
yielding a magnitude limit (1 $\sigma$) of 
 22.6 (FUV) and 22.8 (NUV), in the AB magnitude system, while the AIS has typical exposure
time of about 100 sec, corresponding to limiting magnitudes
  FUV $\approx$ 20. and NUV $\approx$ 20.8.
More relevant to the analysis that will follow is the number of 
sources within each survey with photometric errors better than
specific limits in any band. These numbers are compiled in Table \ref{tlimits},
and can be compared to the total number of sources in these surveys,
given in Table 1 of Seibert et al (2004). Table 1 shows that the 
GALEX MIS survey in the NUV band is somehwat deeper than the SDSS imaging
survey, as is the GALEX Nearby Galaxies Surveys (NGC) that has a similar
exposure time (Bianchi et al. 2004).

For a description of the matching procedure between GALEX and 
SDSS sources, see Seibert et al. (2004). 
Because of the different spatial resolutions, 4.5/6$''$ (GALEX
FUV/NUV) and 1-2$''$ (SDSS), some GALEX sources have more than
one optical counterpart. We excluded from our analysis the sources with
multiple matches (even when the second match was farther away and
the first match is probably significant), because 
the source colors may not be meaningful, due to potential
contamination from other sources unresolved by GALEX.
This restriction excludes about 17\% (MIS) and 18\% (AIS)
 of the sources from the analysis. This factor must be taken into account when
considering the numerical population density (i.e. sources per square
degree) and when comparing with other catalogs,  and other papers in this
volume. 

The accuracy of the GALEX calibration is described by Morrissey et al
(2004). The magnitudes of the IR0.2 release are
tied to the ground-based calibration. The calibration is currently being
improved with on-going observations of standard stars, and appears
to be accurate within 0.1 mag.   Taking into  consideration 
several factors, we restricted our sample for the analysis to
sources with 
photometric errors (excluding the zero-point error)
better than 0.1/0.15 mag (GALEX NUV/FUV bands)
and better than 0.05 mag (SDSS bands). 
We also consider a less stringent error cut
(0.2 mag in any band) for consistency with Seibert et al. (2004),
and give results for both error cuts.

\section{The models}
\label{smodels}

We computed model colors for different astrophysical objects,
in the GALEX and SDSS bands, by using the transmission curves of
the GALEX and SDSS filters available from  the web sites of the projects. 
For normal stars, we used fully blanketed Kurucz models, with \Teff ~ 
from 50000~K to 3000~K, from the grids of Bianchi \& Garcia (in preparation)
and Lejeune et al. (2002). Only models for log g = 5.0 are shown in 
figure 1, to avoid crowding. A  WD model for \Teff =135,000~K, computed
with the TLUSTY (Hubeny \& Lanz 1995) code,  is
also shown.  WD models for lower \Teff ~ overlap with the hottest
main sequence star colors, thus are not shown. 

Colors for integrated populations were computed from Bruzual \& Charlot (2003)
models, for the two extreme cases of Single-burst Star Formation (SSP) and
Continuous Star Formation (CSP). We used both Salpeter,  and Chabrier (2003)
Initial Mass Functions (both cases are plotted, but are almost indistinguishable),
with a range of initial masses 0.1-100\Msun. Only 
 solar metallicity models are shown in the plots. 

For QSO spectra, we use a modified composite spectrum of Francis
et al. (1991). 
At restframe wavelengths shortward of 1000 \AA, the power law is
 $f_\nu$ $\propto$ $\nu^{-1.8}$ (Zheng et al. 1997). The absorption effect of
$Ly\alpha$
absorption follows the work of Press \& Rybicki (1993). 

Models were also reddened with different types of extinction (only MW-type
extinction is shown in the plots). 
The reddening arrows for the different model points are not exactly parallel,
because we reddened the model spectra and then applied the filter transmission
curves, which is more accurate than applying an extinction value appropriate
to the $\lambda$$_{eff}$ of the filter. 

The extensive grids of 
 model colors will be published elsewhere
(Bianchi et al., in prep). 

\section{Analysis.  Results and conclusions}
\label{snalaysis}

\begin{figure*}
\plotone{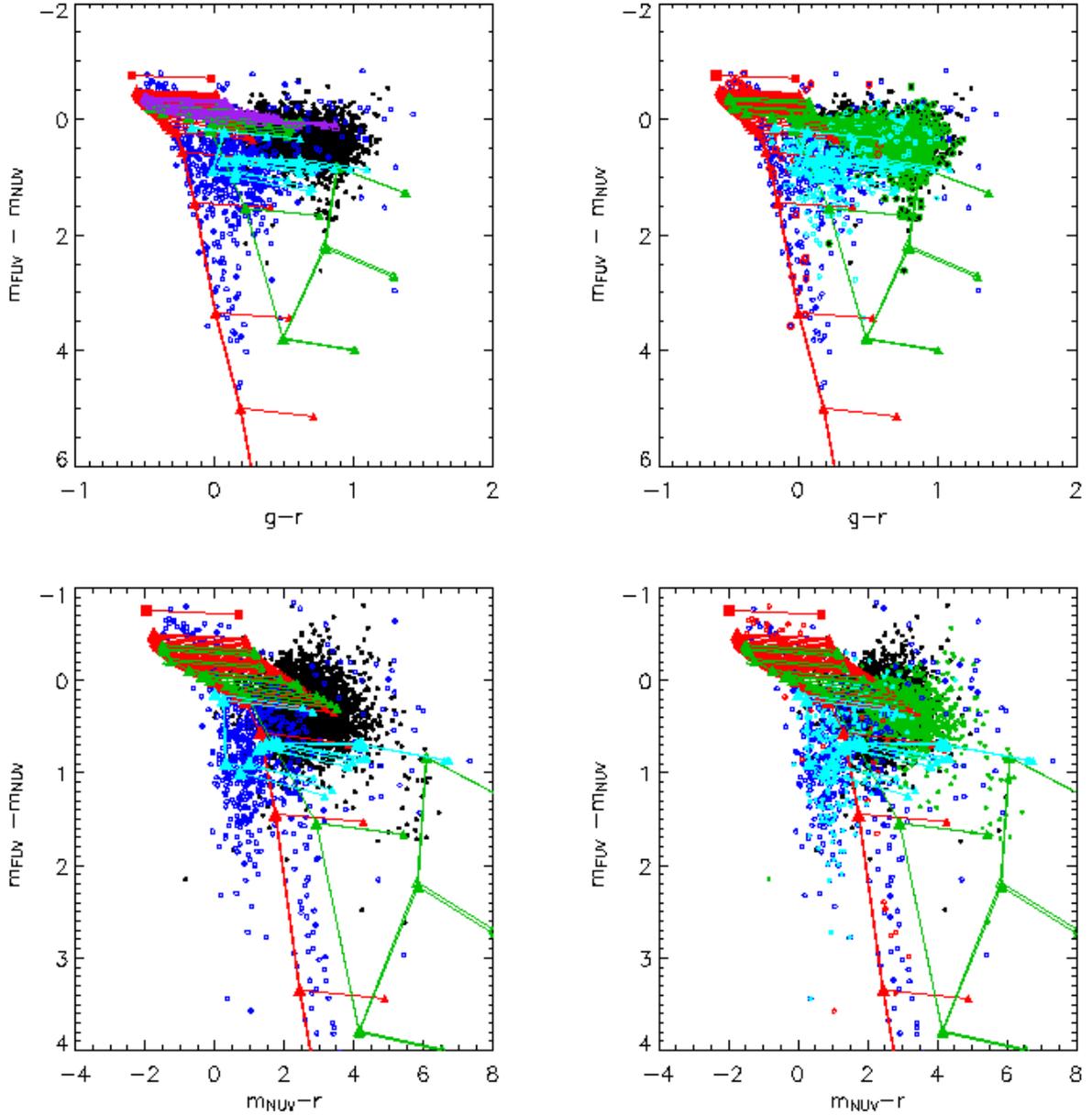} 
\caption{Two color-color diagrams for GALEX IR0.2 MIS sources matched
to the SDSS DR1 sources. Sources classified as point-like
in the SDSS catalog (SDSS type: ``star'') are the blue circles,
sources extended in the SDSS  (SDSS type: ``galaxy'') are plotted as black points.
 On the right-hand panels, the objects with
{\it spectroscopic} classification from the SDSS survey are shown in red
(stars), turquoise (QSOs) and green (galaxies).
Model colors are indicated as: red (dark) triangles: stellar models, from \Teff=50000K
(uppermost point) to 3000K, red square: WD with \Teff=135,000K; 
green triangles: integrated stellar population (Single Star Burst, SSP)
models, from age 0.5Myr to 10 Gyr, and purple triangles: 
Continuous Star Formation (CSP) models for the same ages (shown only in the top left
panel); turquoise
triangles: QSOs in the redshift range 0-4.  Reddening arrows corresponding
to E(B-V)=0.5 (MW reddening, R$_v$=3.1) are shown for each model point.  
 Top panels: NUV-FUV vs g-r.  
It is evident that the classification
star/galaxy based on the source PSF is roughly a good division, however
QSO (mostly point-like) fall in the ``star'' class.
The top (right) panel shows the location of spectroscopically confirmed 
galaxies to be consistent with either reddened young objects, or very old
galaxies where the color becomes again blue due to the HB stars (UV upturn).
The lower-right panel shows that the combination of the UV bands with an optical
band removes the ambiguity.  The other advantage shown by the lower
panels is that QSOs in the redshift range 0-2 can be separated from other
classes of objects by this color combination. 
\label{f_qsoclass} }
\end{figure*}

In Figure \ref{f_qsoclass} we show two examples of color combinations,
that illustrate three different advantages of using UV plus optical bands.
First, the UV colors of starburst (SSP) populations  younger and older than approximately
10$^9$ yrs are ambiguous (as age indicators) because of the 
UV upturn of the old population.
The age ambiguity is largely removed when using an optical band, e.g. V,
as shown in our example by the NUV-r color. 

In the spectro-photometric determination of ages
(and intrinsic luminosities, thus masses) of SSP populations (stellar
clusters, or starburst galaxies), extinction must be determined concurrently
with the physical parameters of the source, mainly the age, since both 
age and extinction affect the observed slope.  
Similarly, one can derive concurrently \Teff ~ and extinction from 
the photometric SED of stars (e.g. Bianchi et al. 2001). 
The second advantage
clearly visible in all panels of Figure \ref{f_qsoclass} is that, while 
fluxes at UV wavelengths suffer by the selective extinction more than optical fluxes,
the {\it color} FUV-NUV is essentially reddening free (for a typical Galactic
extinction law with R$_v$=3.1, as used in Figure \ref{f_qsoclass}, and for 
moderate amounts of reddening). Therefore, the age of an SSP population 
can be determined from this color independently from the estimate of reddening
(see also Bianchi et al 2004), once the ambiguity between young
and old ages is resolved by the additional optical band.   

The third advantage shown by Figure \ref{f_qsoclass} (lower panels)
is that QSOs at low
redshift ($<$2) can be separated from the stars by  their color. This is discussed in
more detail in the next section.
Also, stars of certain types can be separated, and easily identified, by
different color combinations (see also Bianchi \& Martin 1998).

\subsection{Low-Redshift QSO Candidates}

In the lower panels of Figure \ref{f_qsoclass}
we separated the point-like sources which lie on the left-hand side
of the envelope defined by the stellar model colors. 
In this region lie the QSO model colors for low redshift, and about
half of the spectroscopically confirmed QSO SDSS sample.  
 With our stringent error cuts specified
above, we find 
548 
QSO candidates in the MIS sample. 
 Of these, 221  
are spectroscopically  confirmed
QSOs by the SDSS survey. Scaling this number by the  
SDSS QSO confirmation rate, 75\% at low z (Richards et al. 2002) the number
of SDSS ``candidates'' in the same color space becomes 295. 
If we relax the error cut to 0.2 mag in all bands (for comparison with Seibert et al. 2004,
this issue) the numbers become: 1736 
 low redshift QSO GALEX candidates, 263 of which 
spectroscopically confirmed.

For the AIS, which covers at shallower flux limits 92 square degrees,
we find, with the same stringent error limits, 99 low redshit QSO candidates,
of which 45 are included in the SDSS spectroscopically confirmed sample.
However, these error cuts limit the number of sources to about 900, out of
the total almost 100,000.  Using error cuts of 0.2 mag for all bands,
we retain about 2100 AIS sources, and  the number of
low redshift QSO candidates becomes 222, of which 
91 are spectroscopically confirmed.

 Our color-selected sample therefore
 increases the number
of QSO candidates in the redshift range z $<$ 2.   
However, we stress that our photometric selection of objects, defined as the
points outside (left of) the models for single stars in the color-color plot,
may also include binary stars. 
  The distribution of
redshifts within the  spectroscopically confirmed SDSS sample is shown by
Seibert et al. (2004, this issue). 

In Figure 2 we show the characteristics
of the GALEX-selected low redshift QSO-candidate samples versus the 
spectroscopically confirmed subset. 
To evaluate the significance of our candidate sample, we plotted in
Figure 2 (left, dotted line) also the subsample of QSOs candidates brighter than 19.1 in the 
{\it i}-band, which is the completeness limit of the 
low-z QSO (z$<$3) SDSS spectroscopic
 sample, according to Richards et al. (2002).
The histograms in Figure 2 show that  
{\it (i)}  the GALEX-AIS photometrically  selected sample overlaps the 
magnitude range of the SDSS spectroscopic QSO sample,  and is 
numerically larger  by less than a factor of two,  taking into account
the SDSS confirmation rate, {\it (ii)}  the MIS photometrically selected
sample significantly extends to lower magnitudes. In the range of magnitude
overlap with the SDSS spectroscopic survey, the GALEX-MIS selected sample 
is still more numerous by 50\% or less.
In conclusion, our photometric selection of low redshift QSO candidates is 
significantly extending the number of candidates to fainter magnitudes,
in comparison to the SDSS spectrscopic sample, and slightly increases the number of
bright candidates.

The most prominent spectral feature in QSOs' spectra is the flux break due to 
the accumulated Ly$\alpha$ forest absorption. At redshift $z<2$, this feature is 
shortward of the SDSS range, and SDSS uses the general power-law shape at longer 
wavelengths to search for QSOs, without the Ly$\alpha$ break. GALEX has the 
advantage of detecting this feature in the UV bands, thus complements the SDSS
in its quasar search.
With a large number of new QSO candidates, GALEX demonstrates the potential of 
finding many more, fainter quasars, improving our measurement of 
 the QSO luminosity function at $z<2$.

\subsection{The Stars in the MW Halo}

The numerical advantage of spectro-photometric classification over the
SDSS spectroscopic sample  is  greater for stellar
objects (Figure 1).  An additional result is the derivation of
 extinction maps in the MW, which will be the subject of a future paper.
 For instance, in the MIS IR0.2 sample, we have about 2000 stellar objects
(errors $<$ 0.2 mag) in 75 square degrees, i.e. about 27 per square
degree. In a forthcoming paper we will use seven bands (GALEX plus SDSS)
to derive concurrently extinction and \Teff~ by comparison to model colors,
following a modification of the method by Bianchi et al (2001), implemented
by Bianchi and Tolea (2004, in preparation) for HST photometry. 
Very hot stellar objects are, as expected, easily identified by the UV
colors (Figure 1). 
Seibert et al. (2004) present a
 preliminary discussion of specific classes of stellar objects that emerge from 
GALEX plus SDSS  photometric diagrams.

\begin{acknowledgments}
Acknowledgement: 

GALEX (Galaxy Evolution Explorer) is a NASA Small Explorer, launched in April 2003.
We gratefully acknowledge NASA's support for construction, operation,
and science analysis of the GALEX mission,
developed in cooperation with the Centre National d'Etudes Spatiales
of France and the Korean Ministry of 
Science and Technology.  LB is very grateful to J. Maiz, M. Garcia, J. Herald 
for help in constructing  the stellar models,  and to S. Yi and T.Budavari for helpful
discussions.  
\end{acknowledgments}

\newpage

\renewcommand{\arraystretch}{.6}
{\footnotesize
\begin{deluxetable}{lccc}
\tablewidth{0pt}
\tablecaption{The sample of sources in the different bands {\label{tlimits}} }
\tablehead{
\multicolumn{1}{l}{Band}    & \multicolumn{1}{c}{No. Sources}    & 
\multicolumn{1}{c}{No. Sources}   & 
\multicolumn{1}{c}{No. Sources}    \\ 
\multicolumn{1}{l}{ }            &    \multicolumn{1}{c}{(error$<$0.2 mag)} 
                     & \multicolumn{1}{c}{(error$<$0.1 mag)}  & 
                       \multicolumn{1}{c}{(error$<$0.05 mag)}   
}
\startdata
\multicolumn{3}{c}{\it MIS Total Sources\rm } \\
\fuvmag &40355 &  8181 & 1527 \\
\nuvmag &229452& 127253&27905\\
u       &53747 &41468 &32947\\
g       &143432&90350 & 62221\\
r       &169519&108808& 70245\\
i       &144014&91459 &  61190\\
z       &73075 &50203 &  38709\\
\multicolumn{3}{c}{\it AIS Total Sources\rm } \\
\fuvmag & 3835 & 572 & 113  \\
\nuvmag & 32129& 7291&1041\\
u       & 42064&33714&28288\\
g       & 74523& 62889&48277\\
r       & 76750&66673&51821\\
i       & 72309&59527&45681\\
z       & 50227&38400&30508\\
\enddata
\end{deluxetable}
}
\normalsize

\newpage

\newpage

\begin{figure}
\plotone{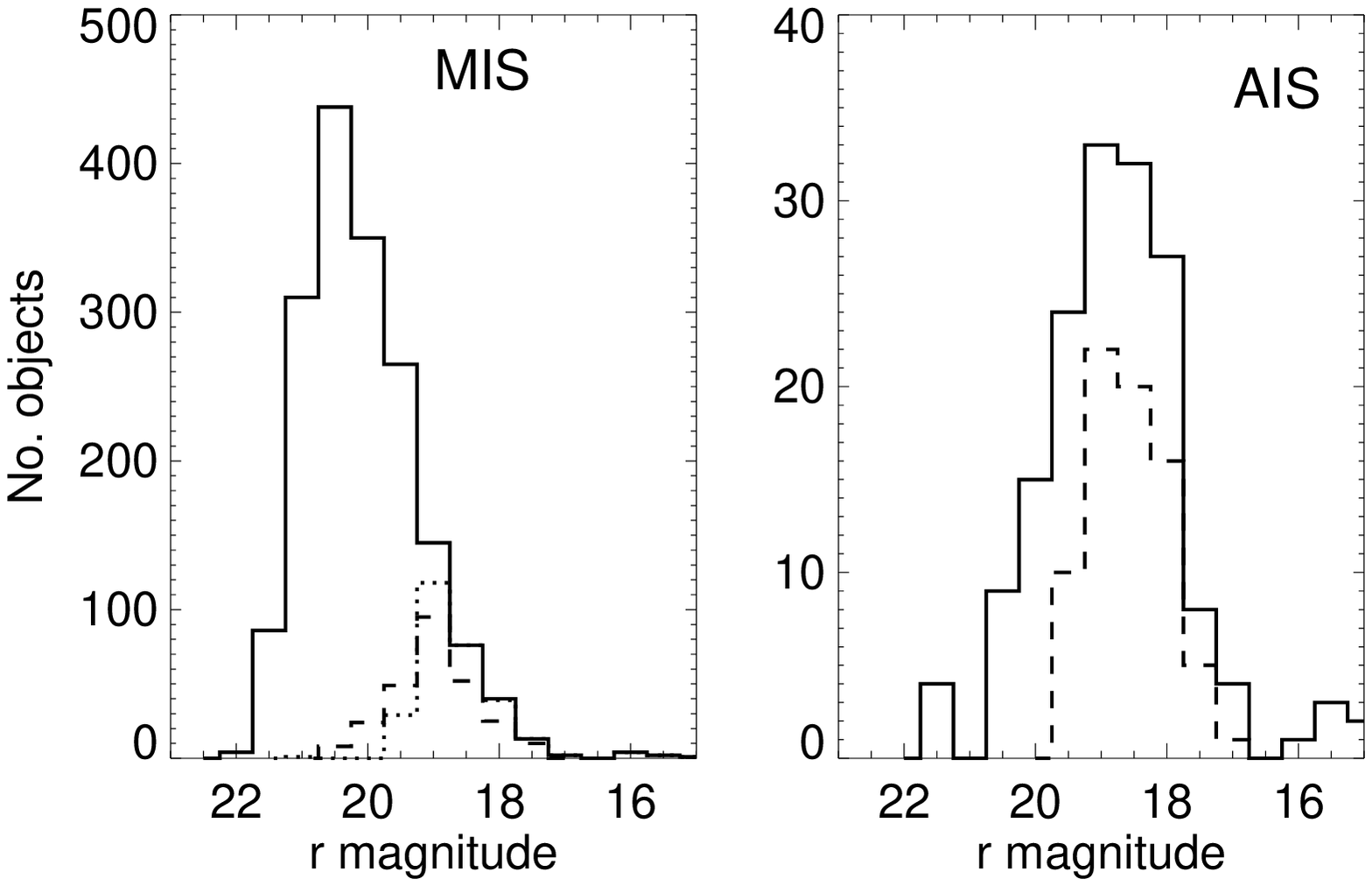} 
~\\
\caption{
Histograms (r-band magnitude) 
of low redshift QSO candidates, selected photometrically from 
 from the MIS (left) and AIS (right) IR0.2 surveys.
The dashed histograms are the subset of the selected samples that are 
spectroscopically confirmed in the SDSS survey. The dotted histogram is the
subset of the MIS sample with {\it i} $<$19.1 mag (see text).
}
\label{f_qsohist}
\end{figure}



\begin{references}
\reference{}Arnouts, S., et al. 2004,  ApJL, this issue
\reference{}Bianchi, L., et al, 2004, ApJL, this issue
\reference{NGS} Bianchi, L., Madore, B., Thilker, D., Gil de Paz, A., and Martin, C., 2004a, in "The Local Group as an
Astrophysical Laboratory", in press 
\reference{NGSAAS}Bianchi, L., Madore, B., Thilker, D., Gil de Paz, A., and the GALEX team, 2004b, AAS 203, 91.12 
\reference{}Bianchi, L., and Garcia, M., 2004, in preparation 
\reference{}Bianchi, L., Scuderi, S., Massey, P., and Romaniello, M. 2001, AJ, 121, 2020 
\reference{}Bianchi, L., \& Martin, C., 1998, in ``UV Astrophysics beyond the IUE Final Archive'', eds. R. Gonzalez-Riestra, W.Wamsteker
and R.A. Harris, ESA SP-413, p. 797
\reference{BC03}Bruzual, G. \& Charlot, G., 2003, MNRAS, 344, 1000
\reference{}Buat, L., et al, 2004, ApJL, this issue
\reference{}Budavari, L., et al, 2004, ApJL, this issue
\reference{}Chabrier, G., 2003, PASP, 115, 763
\reference{}  Francis, P.J., et al. 1991, ApJ, 373, 465.
\reference{} Hubeny, I. \& Lanz, T. 1995, ApJ, 493, 875
\reference{}Lejeune, T., Cuisinier, F., \& Buser, R.\ 1997, \aaps, 125, 229
\reference{}Martin, D. C., et al. 2004, \apj , present volume
\reference{}Morrissey, P., et al. 2004,  \apj ,  present volume
\reference{}Press, W.H. \& Rybicki, G.B. 1993, ApJ, 418, 585
\reference{}Richards, et al. 2002 AJ 123, 2945
\reference{}Rich, M.  et al. 2004, ApJL, this issue
\reference{}Salim, S. et al. 2004, ApJL, this issue
\reference{}Seibert, M. et al. 2004, ApJL, this issue
\reference{}Schlegel, D.~J., Finkbeiner, D.~P., \& Davis, M.\ 1998, \apj, 500, 525 
\reference{}Yi, S. et al, 2004, ApJL, this issue
\reference{}Zheng, W.  et al., 1997, ApJ, 475, 469
\end{references}
\end{document}